\documentclass[aps,preprint,11pt]{revtex4}
\usepackage{amsmath,amssymb,amsthm,color}
\usepackage{epsfig}
\usepackage{graphicx}
\usepackage{amssymb}
\usepackage{amsfonts}
\usepackage{amsmath}
\usepackage{setspace}
\begin{document}
\begin{spacing}{1.5}
\title{Spin entanglement of two spin-$\frac{1}{2}$
particles in a classical gravitational field}
\author{B. Nasr Esfahani }\email{ba_nasre@sci.ui.ac.ir}\affiliation{Department of Physics, Faculty of
Sciences, University of Isfahan , Isfahan 81744, Iran}
\begin{abstract}
The effect of a classical gravitational field on the spin
entanglement of a system of two spin-$\frac{1}{2}$ particles
moving in the curved spacetime is discussed. The system is
described by a two-particle Gaussian wave packet represented in
the momentum space and both the acceleration of the system and the
curvature of the spacetime cause to produce a Wigner rotation
acting on the wave packet as it moves along a path in the curved
spacetime. By calculating the reduced density operator at a final
point, we focus on the spin entanglement of the system. In a
spherically symmetric and static gravitational field, for example
a charged black hole, there can be particular paths  on which the
Wigner rotation is trivial and so the initial reduced density
matrix remains intact. This causes the spin entanglement to be
invariant during the motion. The spin entanglement descends to
zero by increasing the angular velocity of the mean centroid of
the system as well increasing the proper time during which the
centroid moves on its circular path around the center.

\end{abstract}

\maketitle

\section{Introduction\label{sl}}

The field of quantum information has made rapid progress in recent
years. It is important to study all those processes that might
have an effect on quantum entanglement. Special relativistic
effects on quantum entanglement and quantum information is
investigated by many authors
\cite{peres,trashima,alsing,ahn,lee,kim,ging,li}. It is shown that
though Lorentz transformations can change the entanglement of the
spins of massive particles, the total entanglement is invariant
for inertial observers.  Some authors studied the entanglement
between two modes of a free Dirac field as seen by two relatively
accelerated observers \cite{mann,pan}. This entanglement is
degraded by the Unruh effect and asymptotically reaches a
non-vanishing minimum value in the infinite acceleration limit.
Furthermore, using the formalism of quantum fields in curved
spacetimes, extension of the quantum entanglement and
teleportation to the gravitational field in four-dimensional
spacetimes, as well in higher dimensional black hole spacetimes is
discussed \cite{shen,ge}. There exists another approach for
discussing the effect of gravitational field on quantum
information in which, using the concept of local inertial frames,
the special relativistic considerations is extended to general
relativity. Using this approach  and considering a wave packet for
the system, Terashima and Ueda discussed a mechanism of spin
rotation caused by spacetime curvature for spin-$\frac{1}{2}$
particles moving in a gravitational field \cite{terashima3}. It is
shown that this effect gives rise to a spin entropy production
that is unique to general relativity. This means that even if the
spin state of the particle is pure at one spacetime point, it
becomes mixed at another spacetime point. In the same manner, the
gravitational spin entropy production for particles with arbitrary
spin moving in a curved spacetime is discussed \cite{nasr1}.

In the present article, regarding the latter approach, we discuss
the effect of a gravitational field on the spin entanglement of a
system consisting of two spin-$\frac{1}{2}$ particles moving in
the gravitational field. For theoretical reasons, it may be
interesting to quantify the entanglement for such a system. We
describe the system by a two-particle wave packet represented in
the momentum space with a definite width around the momentum of a
mean centroid. We consider the mean centroid for describing the
motion of the system in the gravitational field which is a classic
field in our approach. Thus we require to assume that the
curvature of the gravitational field does not change drastically
within the spacetime scale of the wave packet. Using the concept
of local inertial frames it turns out that both the acceleration
of the centroid and the curvature of the gravitational field cause
to produce a Wigner rotation that affects the wave packet. We
confine the argument to a static spherically symmetric
gravitational field such as a charged black hole which because of
the mathematical form of its metric, allows us to present a
detailed argument. At a given initial point in the curved
spacetime, the momenta of the particles are taken to be separable,
while the spin part is chosen to be entangled as the Bell states.
By calculating the reduced density operator, we focus on the spin
entanglement of the system as moves on a circular path around the
center. In the present problem, one can more closely investigate
how a central massive charged body affects a quantum communication
that takes place around it.

\section{Wigner rotation for two-particle states moving in a gravitational field\label{s2}}

In a curved spacetime the curvature causes to break the global
rotational symmetry. Therefore, the spin of a particle in general
relativity can be defined only locally by switching to an inertial
frame at each point and then invoking the rotational symmetry of
the local inertial frame. A local inertial frame at each point of
a curved spacetime is introduced through a tetrad ${e_a}^\mu(x)$
defined by
\begin{equation}\label{tet}
{e_a}^\mu(x){e_b}^\nu(x)g_{\mu\nu}(x)=\eta_{ab}
\end{equation}
where $g_{\mu\nu}(x)$ is the metric that describes the
gravitational field and $\eta_{ab}=\textrm{diag}(-1,1,1,1)$ is the
Minkowski metric \cite{dinverno}. Greek indices run over the four
general coordinate labels, while Latin indices run over the four
inertial coordinate labels $0,1,2,3$. Now, a spin-$\frac{1}{2}$
particle in a curved spacetime is defined as  a particle whose
states furnish the spin-$\frac{1}{2}$ representation of the local
Lorentz group. A one-particle momentum eigenstate is specified by
$|p,\sigma\rangle$ where $p=(\sqrt{|{\bf p}|^2+m^2c^2},{\bf p})$
is the four-momentum eigenvalue as measured in the local inertial
frame and $\sigma$ denotes the $z$-component of the spin. Then,
the wave packet in the momentum representation for a system of two
non-interacting particles, observed in a local frame located at an
initial point $x^{(i)}$, can be written as
\begin{equation}\label{initial}
     |\Psi^{(i)}\rangle=\sum_{\sigma_1\sigma_2}\int\int d^3\textbf{p}_1\,d^3\textbf{p}_2
     \,\Psi_{\sigma_1\sigma_2}({\textbf{p}}_1,{\textbf{p}}_2)\,|p_1,\sigma_1;p_2,\sigma_2\rangle
\end{equation}
where $p_1$ and $p_2$ are the four-momentum of particles, and
$\Psi_{\sigma_1\sigma_2}(\textbf{p}_1,\textbf{p}_2)$ are wave
functions determining  momentum and spin distribution, normalized
as
\begin{equation}\label{normal}
      \sum_{\sigma_1\sigma_2}\int\int d^3\textbf{p}_1\,d^3\textbf{p}_2\,
      |\Psi_{\sigma_1\sigma_2}(\textbf{p}_1,\textbf{p}_2)|^2=1
\end{equation}
provided that $\langle
p'_1,\sigma'_1;p'_2,\sigma'_2|p_1,\sigma_1;p_2,\sigma_2\rangle =
\delta^3(\textbf{p}'_1-\textbf{p}_1)\,\delta^3(\textbf{p}'_2-\textbf{p}_2)\,
\delta_{\sigma'_1\sigma_1}\,\delta_{\sigma'_2\sigma_2}$. We can
use $\Psi_{\sigma_1\sigma_2}(\textbf{p}_1,\textbf{p}_2)$ for
specifying the momentum entanglement, the spin entanglement or
even the entanglement between spins and momenta.

It is required to assume that the spacetime curvature does not
change drastically within the spacetime scale of the wave packet.
Otherwise, we encounter some problems in a classical approach for
the gravitational field. Suppose that the two-particle wave packet
(\ref{initial}) has a mean centroid that in a semiclassical
approach one can regard it to describe the motion of the center of
mass of the two-particle system. The momentum of each particle is
assumed to be distributed properly around the momentum of the
centroid. Assume that the centroid moves along a specified path
$x^\mu(\tau)$ in the curved spacetime. So, the four-momentum of
the centroid as measured in a local inertial frame will be as
$q^a(x)={e^a}_{\mu}(x)(m\,dx^\mu/d\tau)$. Suppose that at $\tau_i$
the centroid  locates at an initial point $x_i=x^\mu(\tau_i)$.
Then, the centroid moves along the path $x^\mu(\tau)$ and at a
proper time $\tau_f$ it reaches to a final point
$x_f=x^\mu(\tau_f)$. Consider two adjacent points on the path such
that $d\tau$ be the infinitesimal proper time between them and at
each point  a local inertial frame is located. It is shown
\cite{terashima2} that these frames are related by a local Lorentz
transformation
${\Lambda^a}_b(x)={\delta^a}_b+{\lambda^a}_b(x)d\tau$, where
\begin{equation}\label{lamb}
    {\lambda^a}_b(x)=-\frac{1}{mc^2}\left[a^a(x)q_b(x)-q^a(x)a_b(x)\right]
    +u^\mu(x)[{e_b}^\nu(x)\nabla_\mu{e^a}_\nu(x)]
\end{equation}
where $u^\mu(x)$ is the four-velocity of the centroid and
$a^a(x)={e^a}_\mu(x)[u^\nu(x)\nabla_\nu u^\mu(x)]$ is its
four-acceleration as measured in the local frame. A classical
force is needed to produce this acceleration, so for geodesic
motions that $a^a(x)=0$, no force is needed. The first part of
(\ref{lamb}) is accelerated related, existing even in special
relativity, and the second part is curvature related, that arises
from the change in the local inertial frame along the path. One
can construct a global Lorentz transformation $\Lambda(x_f,x_i)$
by adding the successive local Lorentz transformations as
\cite{terashima2}
\begin{equation}\label{glt}
      \Lambda(x_f,x_i)=\textrm{T}\exp\left[\int_{x_i}^{x_f}
      \lambda(x(\tau))d\tau\right],
\end{equation}
where $\textrm{T}$ is the time-ordering operator and $\lambda(x)$
is a matrix  whose elements are given by (\ref{lamb}).

Now imagine that when the centroid is at the initial point $x_i$,
the corresponding wave packet as viewed from a local inertial
frame located at $x_i$ is given by (\ref{initial}). When the
centroid reaches to the final point $x_f$, the corresponding wave
packet, as viewed from a local inertial frame located at that
point, can be written as
\begin{eqnarray}\label{final}
      |\Psi^{(f)}\rangle&=&U(\Lambda_1(x_f,x_i))\otimes U(\Lambda_2(x_f,x_i))|\Psi^{(i)}\rangle\\\nonumber
       &=&\sum_{\sigma_1\sigma_2}\int d^3\textbf{p}_1\sqrt{\frac{(\Lambda_1 p_1)^0}{p_1^0}}\int d^3
        \textbf{p}_2\sqrt{\frac{(\Lambda_2p_2)^0}{p_2^0}}\Psi_{\sigma_1\sigma_2}(\textbf{p}_1,\textbf{p}_2)\\\nonumber
       &\times& \sum_{\sigma'_1\sigma'_2}D_{\sigma'_1\sigma_1}(W(\Lambda_1,p_1))D_{\sigma'_2\sigma_2}(W(\Lambda_2,p_2))
      |\Lambda_1 p_1,\sigma'_1;\Lambda_2p_2,\sigma'_2\rangle\,,
\end{eqnarray}
where $U(\Lambda(x_f,x_i))$ is a unitary operator,  $W(\Lambda,p)$
is the Wigner rotation operator corresponding to
$\Lambda(x_f,x_i)$ and $D_{\sigma'\sigma}$ denotes the
two-dimensional representation of the Wigner rotation operator
\cite{weinberg}. The Wigner rotation operator can be expressed
explicitly as \cite{terashima3}
\begin{equation}\label{Wig}
      W(\Lambda(x_f,x_i),p)=\textrm{T}
     \exp\left[\int_{x_i}^{x_f}w(x(\tau))d\tau\right],
\end{equation}
where $w$ is a matrix whose elements are given by
\begin{equation}\label{dabelu}
      {w^i}_j(x)={\lambda^i}_j(x)+\frac{{\lambda^i}_0(x)p_j(x)
      -\lambda_{j0}(x)p^i(x)}{p^0(x)+mc},
\end{equation}
where $i$ and $k$ run over the three spatial inertial frame labels
$(1,2,3)$. Note that $w_{ii}=0$.

In the following, as the background gravitational field, we
consider a static spherically symmetric spacetime described with
the metric
\begin{equation}\label{metric}
     ds^2=-c^2e^{2A(r)}dt^2+e^{2B(r)}dr^2+
     r^2\left(d{\theta}^2+\sin^2\theta\,d\phi^2\right).
\end{equation}
The asymptotic flatness assumption imposes the following
conditions on $A(r)$ and $B(r)$,
\begin{equation}
      \lim_{r\rightarrow\infty}A(r)=
      \lim_{r\rightarrow\infty}B(r)=0.
\end{equation}
We require to introduce a local inertial frame at each point of
the spacetime. Hence we employ the tetrad
\begin{equation}\label{dstet}
      {e_0}^t=\frac{1}{c}e^{-A(r)},\,\,\,\,\,\,\,\,\,\,{e_1}^r=e^{-B(r)},\,\,\,\,\,\,\,\,\,\,
      {e_2}^{\theta}=\frac{1}{r},\,\,\,\,\,\,\,\,\,\,{e_3}^{\phi}=\frac{1}{r\sin\theta},
\end{equation}
with all the other components being zero.

Generally, in a given gravitational field, the mathematical form
of the local Lorentz transformation and the corresponding Wigner
rotation depends on the path along which the centroid (wave
packet) moves. For the metric (\ref{metric}), regarding its
spherical symmetry, we choose conveniently the equatorial plane
$\theta=\frac{\pi}{2}$ as the plane of motion and suppose that the
centroid is moving with a constant speed $v$ on a specified circle
of radius $r$ around the center. This motion is generally a
non-geodesic motion and so a central force is needed to maintain
the system on the orbit. Then, the components of the four-momentum
of the centroid in the local inertial frame are
\begin{equation}\label{4moment}
      q^0=\gamma mc\,\,\,\,\,\,\,\,\,\,\,\,\,\,\,\,q^1=q^2=0.
      \,\,\,\,\,\,\,\,\,\,\,\,\,\,\,q^3=\gamma mv.
\end{equation}
In this case, the acceleration has only one non-zero component as
\begin{equation}
      a^1(r)=c^2\gamma^2e^{-B(r)}\left(A'(r)-\frac{1}{r}\frac{\gamma^2-1}{\gamma^2}\right),
\end{equation} and then the matrix $\lambda$ has only four non-zero
elements as
\begin{eqnarray}
      {\lambda^1}_0(r)={\lambda^0}_1(r)=c\,\gamma(\gamma^2-1)e^{-B(r)}\left[A'(r)-\frac{1}{r}\right],\\
      {\lambda^1}_3(r)=-{\lambda^3}_1(r)=-\gamma^3ve^{-B(r)}\left[A'(r)-\frac{1}{r}\right].
\end{eqnarray}
These components as substituted in (\ref{dabelu}), lead to the
following non-zero time-independent elements for the local Wigner
rotation matrix $w$
\begin{eqnarray}\label{16}
      {w^1}_3=-{w^3}_1&=&-c\,e^{-B(r)}\left[A'(r)-\frac{1}{r}\right]\\\nonumber
       &&\times \frac{q^3}{mc}\sqrt{\left(\frac{q^3}{mc}\right)^2+1}
       \left(\sqrt{\left(\frac{q^3}{mc}\right)^2+1}-\frac{q\,p^3}{\sqrt{\textbf{p}^2+m^2c^2}+mc}\right).
\end{eqnarray}
Consequently, for circular motions in a general spherically
symmetric spacetime, the Wigner rotation operator (\ref{Wig})
reduces to
\begin{equation}\label{wigop}
     W(\Lambda(\tau))=\frac{1}{2}\left(
       \begin{array}{ccc}
          \cos\Theta &0&\sin\Theta\\
            0&1&0\\
            -\sin\Theta&0& \cos\Theta\\
       \end{array}
      \right).
\end{equation}
where $\Theta={w^1}_3\tau$ with $\tau=\tau_f-\tau_s$. This
operator shows a rotation about the 2-axis and  has a
2-dimensional representation as
\begin{equation}\label{D}
      D(\Theta)=e^{-iJ_2\Theta}=\left(
       \begin{array}{cc}
      \cos \frac{\Theta}{2}&-\sin \frac{\Theta}{2}\\
      \sin \frac{\Theta}{2}&\cos \frac{\Theta}{2}
       \end{array}
      \right),
\end{equation}
where $J_2$ is the 2-component of the angular momentum operator.

One may argue that if $A'(r)=1/r$ identically, $\Theta$ will
vanish and there is no Wigner rotation. However, this condition
leads to a logarithmic $r$-dependence for $A(r)$, which
contradicts the asymptotic flatness assumption we made for the
metric (\ref{metric}). Of course, the condition can be fulfilled
at some distinct radii.

\section{Reduced density matrix}

We intend to calculate the entanglement between the spins of
particles described by the final state (\ref{final}). To do this
we refer to the corresponding reduced density operator
$\varrho^{(f)}$ obtained by taking the trace over the momentum of
the density operator
$\rho^f=|\Psi^{(f)}\rangle\langle\Psi^{(f)}|$, that is
\begin{eqnarray}\label{varrho}\nonumber
      \varrho^{(f)}_{\sigma'_1\sigma'_2,\sigma_1\sigma_2}&=&\sum_{\sigma''_1\sigma''_2}
       \sum_{\sigma'''_1\sigma'''_2}\int\int
       d^3\textbf{p}_1\,d^3\textbf{p}_2\\
        &\times&\left[D_{\sigma'_1\sigma''_1}(W(\Lambda_1,p_1))
       \Psi_{\sigma''_1\sigma''_2}(\textbf{p}_1,\textbf{p}_2)
       D_{\sigma'_2\sigma''_2}(W(\Lambda_2,p_2))\right]\\\nonumber
       &\times&\left[D_{\sigma_1\sigma'''_1}(W(\Lambda_1,p_1))
       \Psi_{\sigma'''_1\sigma'''_2}(\textbf{p}_1,\textbf{p}_2)
       D_{\sigma_2\sigma'''_2}(W(\Lambda_2,p_2))\right]^*.
\end{eqnarray}
To follow the argument we need to choose an explicit form for
$\Psi_{\sigma_1\sigma_2}(\textbf{p}_1,\textbf{p}_2)$. We assume no
entanglement between spins and momenta, as well as no entanglement
between momenta, however, we assume a maximum spin entanglement by
choosing the spin part to be one of the Bell states.  Therefore,
we can make a column vector as
\begin{equation}\label{mateh}
      \Psi(\textbf{p}_1,\textbf{p}_2)=f(\textbf{p}_1)f(\textbf{p}_2)\chi_i
\end{equation}
where $f(\textbf{p})$ is a normalized function and $\chi_i$ is is
one of the Bell states
\begin{eqnarray}\label{bell}\nonumber
     \chi_1=\frac{1}{\sqrt{2}}\left(
     \begin{array}{c}
      1\\0\\0\\1
      \end{array}
      \right),\hspace{2cm}
     \chi_2=\frac{1}{\sqrt{2}}\left(
      \begin{array}{c}
     1\\0\\0\\-1
      \end{array}
     \right),\\
     \chi_3=\frac{1}{\sqrt{2}}\left(
      \begin{array}{c}
     0\\1\\1\\0
      \end{array}
     \right),\hspace{2cm}
     \chi_4=\frac{1}{\sqrt{2}}\left(
     \begin{array}{c}
     0\\1\\-1\\0
     \end{array}
     \right).
\end{eqnarray}
Substituting the components of (\ref{mateh}) in (\ref{varrho}), we
obtain the final reduced density matrix components as
\begin{eqnarray}\label{frd12}
      \varrho^{(f)}_{\sigma'_1\sigma'_2,\sigma_1\sigma_2}&=&\frac{1}{2}\int d^3\textbf{p}_1|f(\textbf{p}_1)|^2\int d^3
       \textbf{p}_2|f(\textbf{p}_2)|^2\\\nonumber
       && \times[\,\,D_{\sigma'_1\uparrow}(\Theta_1)D_{\sigma'_2\uparrow}(\Theta_2)D^*_{\sigma_1\uparrow}(\Theta_1)
        D^*_{\sigma_2\uparrow}(\Theta_2)\\\nonumber
        &&\,\,\,\,\,\,\,\,\,\pm\, D_{\sigma'_1\uparrow}(\Theta_1)D_{\sigma'_2\uparrow}(\Theta_2)
        D^*_{\sigma_1\downarrow}(\Theta_1)D^*_{\sigma_2\downarrow}(\Theta_2)\\\nonumber
        &&\,\,\,\,\,\,\,\,\,\pm \,D_{\sigma'_1\downarrow}(\Theta_1)D_{\sigma'_2\downarrow}(\Theta_2)
        D^*_{\sigma_1\uparrow}(\Theta_1)D^*_{\sigma_2\uparrow}(\Theta_2)\\\nonumber
       &&\,\,\,\,\,\,\,\,\,+D_{\sigma'_1\downarrow}(\Theta_1)D_{\sigma'_2\downarrow}(\Theta_2)
      D^*_{\sigma_1\downarrow}(\Theta_1)D^*_{\sigma_2\downarrow}(\Theta_2)\,\,]\nonumber,
\end{eqnarray}
where the upper (lower)  sign corresponds to the choice $\chi_1$
($\chi_2$) , and
\begin{eqnarray}\label{frd34}
      \varrho^{(f)}_{\sigma'_1\sigma'_2,\sigma_1\sigma_2}&=&\frac{1}{2}\int d^3\textbf{p}_1|f(\textbf{p}_1)|^2\int d^3
        \textbf{p}_2|f(\textbf{p}_2)|^2\\\nonumber
        && \times[\,\,D_{\sigma'_1\uparrow}(\Theta_1)D_{\sigma'_2\downarrow}(\Theta_2)D^*_{\sigma_1\uparrow}(\Theta_1)
        D^*_{\sigma_2\downarrow}(\Theta_2)\\\nonumber
         &&\,\,\,\,\,\,\,\,\,\pm\, D_{\sigma'_1\downarrow}(\Theta_1)D_{\sigma'_2\uparrow}(\Theta_2)
         D^*_{\sigma_1\uparrow}(\Theta_1)D^*_{\sigma_2\downarrow}(\Theta_2)\\\nonumber
         &&\,\,\,\,\,\,\,\,\,\pm \,D_{\sigma'_1\uparrow}(\Theta_1)D_{\sigma'_2\downarrow}(\Theta_2)
        D^*_{\sigma_1\downarrow}(\Theta_1)D^*_{\sigma_2\uparrow}(\Theta_2)\\\nonumber
       &&\,\,\,\,\,\,\,\,\,+D_{\sigma'_1\downarrow}(\Theta_1)D_{\sigma'_2\uparrow}(\Theta_2)
       D^*_{\sigma_1\downarrow}(\Theta_1)D^*_{\sigma_2\uparrow}(\Theta_2)\,\,]\nonumber.
\end{eqnarray}
where the upper (lower)  sign corresponds to $\chi_3$ ($\chi_4$).
Here the components of $D(\Theta_1)$ or $D(\Theta_2)$ are given by
(\ref{D}).

We have suggested that the centroid moves along a circle with a
definite momentum, however, from a quantum mechanical point of
view we imagine that the momentum represented wave packet is
distributed properly around the momentum of the centroid. In what
follows, we choose a suitable form for the distribution function
as
\begin{equation}\label{f(p)}
      f(\textbf{p})=\frac{\sqrt{\delta(p^1)}\sqrt{\delta(p^2)}}{\sqrt{\pi^{1/2}\beta\,mc}}
      \exp{\left[-\frac{(p^3-q^3)^2}{2\beta^2 m^2c^2}
      \right]},
\end{equation}
which describes a sharp distribution about $p^1=0$ and $p^2=0$,
and a Gaussian distribution about $p^3=q^3$ with a width specified
by $\beta$.

Let us apply (\ref{f(p)}) in (\ref{frd12}) and (\ref{frd34}) and
define
\begin{equation}\label{C}
       \mathcal{C}=\frac{1}{(\sqrt{\pi})\beta}\int_{-\infty}^\infty
       d\,p\,\,\exp\left[-\,\frac{(p-q)^2}{\beta^2}\right]\cos\Theta,
\end{equation}
and
\begin{equation}\label{S}
      \mathcal{S}=\frac{1}{(\sqrt{\pi})\beta}\int_{-\infty}^\infty
       d\,p\,\,\exp\left[-\,\frac{(p-q)^2}{\beta^2}\right]\sin\Theta,
\end{equation}
where $p=\frac{p^3}{mc}$ and $q=\frac{q^3}{mc}$ are dimensionless
parameters and $\Theta$ is given by
\begin{eqnarray}\label{Theta}
      \Theta &=&-c\,\tau\,e^{-B(r)}\left[A'(r)-\frac{1}{r}\right]\\\nonumber
       &&\times
       q\sqrt{q^2+1}\left(\sqrt{q^2+1}-\frac{q\,p}{\sqrt{p^2+1}+1}\right),
\end{eqnarray}
according to (\ref{16}). Then, after doing some manipulations we
obtain the final reduced density matrices as
\begin{equation}
      \varrho^{(f)}=\frac{1}{4}\left(
      \begin{array}{cccc}
     1\pm\,\mathcal{C}^2\pm\,\mathcal{S}^2&0&0&1
      \pm\,\mathcal{C}^2\pm\,\mathcal{S}^2\\
       0&1\mp\,\mathcal{C}^2\mp\,\mathcal{S}^2&-1\pm\,\mathcal{C}^2\pm\,\mathcal{S}^2&0\\
       0&-1\pm\,\mathcal{C}^2\pm\,\mathcal{S}^2&1\mp\,\mathcal{C}^2\mp\,\mathcal{S}^2&0\\
      1\pm\,\mathcal{C}^2\pm\,\mathcal{S}^2&0&0&1\pm\,\mathcal{C}^2\pm\,\mathcal{S}^2
      \end{array}
      \right),
\end{equation}
where the upper (lower) sign corresponds to $\chi_1$ ($\chi_4$),
and
\begin{equation}
      \varrho^{(f)}=\frac{1}{4}\left(
      \begin{array}{cccc}
      1\pm\,\mathcal{C}^2\mp\,\mathcal{S}^2&
      \mp2\,\mathcal{C}\,\mathcal{S}&\mp\,2\,\mathcal{C}\,\mathcal{S}&-1\mp\,\mathcal{C}^2\pm\,\mathcal{S}^2\\
      \mp\,2\,\mathcal{C}\,\mathcal{S}&1\mp\,\mathcal{C}^2\pm\,\mathcal{S}^2&
      1\mp\,\mathcal{C}^2\pm\,\mathcal{S}^2&\pm\,2\,\mathcal{C}\,\mathcal{S}\\
      \mp\,2\,\mathcal{C}\,\mathcal{S}&1\mp\,\mathcal{C}^2\pm\,\mathcal{S}^2&
      1\mp\,\mathcal{C}^2\pm\,\mathcal{S}^2&\pm\,2\,\mathcal{C}\,\mathcal{S}\\
      -1\mp\,\mathcal{C}^2\pm\,\mathcal{S}^2&\pm\,2\,\mathcal{C}\,\mathcal{S}&\pm\,2\,\mathcal{C}\,\mathcal{S}&
      1\pm\,\mathcal{C}^2\mp\,\mathcal{S}^2
    \end{array}
    \right),
\end{equation}
where the upper (lower) sign is used for $\chi_2$ ($\chi_3$).

It is interesting to consider a situation in which the centroid
falls along a radial geodesic with a velocity $v$, as measured in
a local inertial frame. Then we can write $v=c{e^1}_r dr/{e^0}_t
dt=e^{-A(r)}e^{B(r)}(dr/dt)$ which leads to $q^0=\gamma m c$ and
$q^1=\gamma m v$ as the non-zero components of the four-momentum
of the centroid. Since this is a geodesic motion, the acceleration
part of (\ref{lamb}) vanishes and the curvature part gives only
${\lambda^0}_1={\lambda^1}_0=-c \gamma A'(r)e^{-B(r)}$ which
consists of a boost along the 1-axis. In the present case, all the
components of ${w^i}_j$ are zero as (\ref{dabelu}) implies. Thus,
the Wigner rotation (\ref{Wig}) is trivial and its two-dimensional
representation simply becomes
$D_{\sigma'\sigma}=\delta_{\sigma'\sigma}$, which identically
leads to the result $\varrho^{(f)}=\varrho^{(i)}$ independent of
spin entanglements.  This means, in the case of radial motion the
reduced density matrix is invariant  and there is no spin
decoherence. Of course, there  exists still a global Lorentz
transformation obtained from (\ref{glt}) which transforms the
initial wave packet (\ref{initial}) into a boosted frame along the
1-axis via the operator $U(\Lambda_1(r_f,r_i))\otimes
U(\Lambda_2(r_f,r_i))$.

A possible application of the above result is to suggest a general
relativistic invariant protocol for quantum communication. For an
observer falling on a radial geodesic the information stored in
spins of two entangled spin-$\frac{1}{2}$ particles remains
perfect.

\section{Spin Entanglement}

The entanglement between spins as viewed by an observer located at
a local frame at the final point $x^{(f)}$ is obtained by
calculating the Wootters' concurrence \cite{wootters}, denoted as
$\textsf{C}(\varrho)$. It is defined as
$\textsf{C}(\varrho)=\textrm{max}\{0,\lambda_1-\lambda_2-\lambda_3-\lambda_4\}$
where $\lambda_i$s are the square root of the eigenvalues, in
decreasing order, of non-hermitian matrix $\varrho\tilde{\varrho}$
where
$\tilde{\varrho}=(\sigma_y\otimes\sigma_y)\varrho^*(\sigma_y\otimes\sigma_y)$.
Note that each $\lambda_i$ is a non-negative real number. Then,
the entanglement $\textsf{E}(\varrho)$ can be calculated as
\begin{equation}
       \textsf{E}(\varrho)=h\left(\frac{1+\sqrt{1-\textsf{C}^2}}{2}\right)
\end{equation}
where
\begin{equation}
      h(x)=-x\log_2x-(1-x)\log_2(1-x).
\end{equation}

After doing some manipulations it turns out that the eigenvalues
of $\varrho^{(f)}{\tilde{\varrho}}^{(f)}$ for all of the matrices
(\ref{frd12}) and (\ref{frd34}) are the same and are obtained as
$$\left\{\frac{1}{4}\left(1+\mathcal{C}^2+\mathcal{S}^2\right)^2,
\frac{1}{4}\left(1-\mathcal{C}^2-\mathcal{S}^2\right)^2,0,0\right\}.$$
Then the concurrence in this case becomes
$\textsf{C}=\mathcal{C}^2+\mathcal{S}^2$. This means that the
entanglement for all of the states can be calculated by a unique
relation as
\begin{eqnarray}\nonumber\label{emp}
     \textsf{E}=-\frac{1+\sqrt{1-(\mathcal{C}^2+\mathcal{S}^2)^2}}{2}
      \log_2\left(\frac{1+\sqrt{1-(\mathcal{C}^2+\mathcal{S}^2)^2}}{2}\,\right)\\
     -\frac{1-\sqrt{1-(\mathcal{C}^2+\mathcal{S}^2)^2}}{2}
     \log_2\left(\frac{1-\sqrt{1-(\mathcal{C}^2+\mathcal{S}^2)^2}}{2}\,\right).
\end{eqnarray}

\section{Discussion}

For evaluating these entanglements we need to specify the back
ground gravitational field by determining the unknown functions
$A(r)$ and $B(r)$ in (\ref{Theta}). The properties of these
functions affect directly the behavior of the Wigner rotation in
terms of $r$. In the following we illustrate the argument for a
charged black hole \cite{dinverno}. As we will see below the
particular forms of $A(r)$ and $B(r)$ for this spacetime provides
interesting consequences for the behavior of the entanglement in
terms of $r$. The existence of two parameters corresponding to the
mass and the charge of this black hole, allows us to furnish a
detailed argument on the gravitational spin decoherence. This
seems to be enough to motivate us to consider such a gravitational
field. Therefore, we choose
\begin{equation}\label{cb}
      e^{2A(r)}=e^{-2B(r)}=1-\frac{r_s}{r}+\frac{\varepsilon^2}{r^2},
\end{equation}
where $r_s$ and $\varepsilon$ relate to the mass and the charge of
black hole, respectively. If ${r_s}^2<4\varepsilon^2$ the
expression (\ref{cb}) will have no real roots. Hence, it follows
that the metric (\ref{metric}) is  singular only at the origin
$r=0$. This singularity is intrinsic and in the present case is
called naked singularity, since there is no event horizon. The
more interesting case occurs when ${r_s}^2\geq4\varepsilon^2$,
such that the metric has two (or one) other singularities where
(\ref{cb}) vanishes, namely, at
$r_\pm=(r_s\pm\sqrt{{r_s}^2-4\varepsilon^2}\,)/2$. These roots are
called event horizons. At any event horizon the tetrad
(\ref{dstet}) diverges and consequently the local Lorentz
transformation (\ref{lamb}) diverges. Therefore, the Wigner
rotation angle $\Theta$ in (\ref{Theta}) must diverge at any
horizon.

Substituting (\ref{cb}) in (\ref{Theta}), the angle $\Theta$ can
be written as
\begin{eqnarray}\label{Thetacb}
     \Theta&=&2\pi\left(\frac{\tau}{\tau_s}\right)\left(\frac{2\,z^2-3z+4\xi^2}{2\,z^2\sqrt{z^2-z+\xi^2}}\right)\\\nonumber
     &&\times q\sqrt{q^2+1}\left(\sqrt{q^2+1}
     -\frac{q\,p}{\sqrt{p^2+1}+1}\right),
\end{eqnarray}
where $\tau_s$ is a constant proper time during which a photon
rotates once on a circle of radius $r_s$, and
$\xi^2=\frac{\varepsilon^2}{{r_s}^2}$ and $z=\frac{r}{r_s}$ are
dimensionless parameters.

For $\xi^2<1/4$, we encounter two event horizons located at
$z_\pm=(1\pm\sqrt{1-4\xi^2})/2$. As (\ref{Thetacb}) implies, in
the region between the horizons, that is $z_-<z<z_+$, the angle
$\Theta$ becomes imaginary and our argument fails there. So we
consider only the region $z>z_+$. (The region $z<z_-$ is not
physically interesting in our argument). In this case  $\Theta$
has two zeros however only one of them
$z_1=(3+\sqrt{9-32\xi^2})/4$ is accessible. For $\xi^2=1/4$, only
one event horizon occurs which is located at $z=1/2$ and $\Theta$
has an accessible zero at $z=1$. In the case of naked singularity
$\xi^2>1/4$ and there is no event horizon. Then $\Theta$ is real
everywhere and it is singular only at $z=0$. There exist two
accessible zeros for $\Theta$ at
$z_{1,2}=(3\pm\sqrt{9-32\xi^2})/4$ provided $\xi^2<9/32$,
otherwise it has no zeros.

Substitute $\Theta$ given by (\ref{Thetacb}) in (\ref{C}) and
(\ref{S}),  it turns out that after integrating, the entanglement
$\textsf{E}$ given by (\ref{emp}) depends formally to the
parameters $\xi$, $\tau, \beta, z$ and $q$. There is no analytical
solution for the integrals and so we attend to a numerical
approach for solving them. In the following we indicate the
behavior of $\textsf{E}$ in some graphs all obtained numerically.
\begin{figure}
\begin{center}
\includegraphics[width=8cm,height=8cm,angle=0]{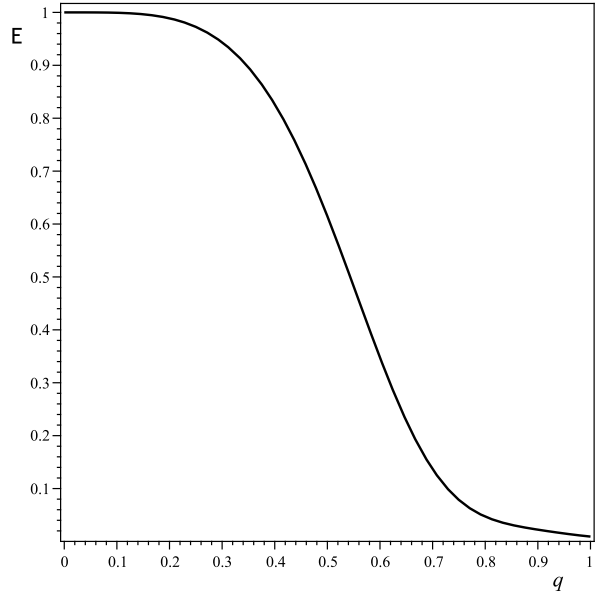}
\end{center}
      \caption{Plot of $\textsf{E}$ for  fixed $\xi$, $z$, $\tau$ and $\beta$ but varying
      $q$. As $q$ increases more spin decoherence occurs and so the entanglement
      descends.}\label{fig1}
\end{figure}

\begin{figure}
\begin{center}
\includegraphics[width=8cm,height=8cm,angle=0]{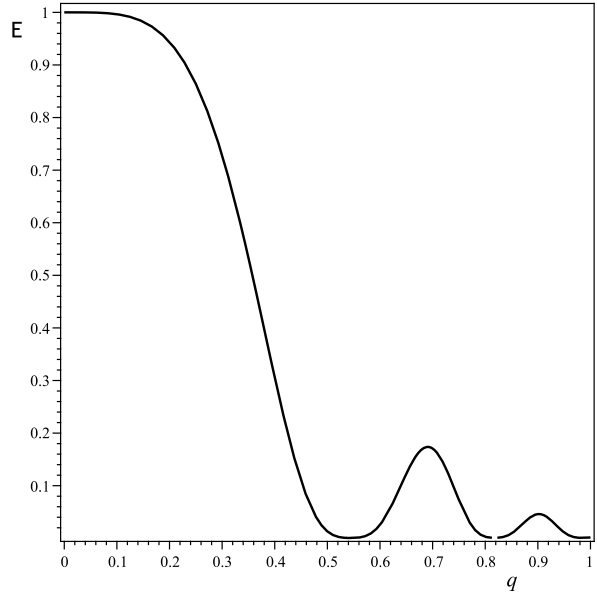}
\end{center}
    \caption{The same plot of figure 1 but for a larger
    value of $\beta$, that is a wider momentum distribution. As $q$
    increases the curve again descends but by performing aperiodic
    oscillations.}\label{fig2}
\end{figure}

In Fig. \ref{fig1},  the entanglement  is sketched versus a $q$
for $\xi^2=0.265$, $\tau=5\tau_0$, $z=1.6$ and $\beta=1$. We see
that by increasing the angular velocity of the centroid on the
circle, $\textsf{E}$ descends uniformly from 1 to an asymptotic
value of zero. Such a behavior is justified because, for a
constant $\tau$, by increasing $q$, the centroid travels more on
its circular trajectory in the gravitational field and so more
spin decoherence occurs. It is remarkable that increasing the
width $\beta$, has a meaningful effect on the behavior of
$\textsf{E}$ in terms of $q$. This point is shown in Fig.
\ref{fig2} which is plotted for $\beta=4$ and the other parameters
as those of Fig. \ref{fig1}. The entanglement again diminishes but
by performing an aperiodic oscillation. This can be justified by
noting that as $\beta$ increases, the exponentials in the
integrals (\ref{C}) and (\ref{S}) approach the unity , so the
cosine or sine terms play a more dominant role.

\begin{figure}
\begin{center}
\includegraphics[width=8cm,height=8cm,angle=0]{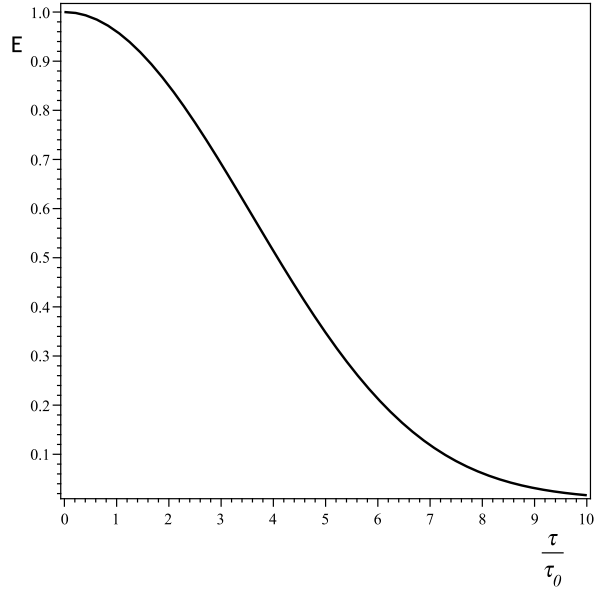}
\end{center}
    \caption{Plot of $\textsf{E}$ for fixed $\xi$, $z$, $q$ and $\beta$ but varying $\tau$.
    Increasing $\tau$ also causes to occur more spin decoherence and so the entanglement
    diminishes.}\label{fig3}
\end{figure}

One may also plot $\textsf{E}$ in terms of $\tau/\tau_0$ where
$\tau$ is the proper time during which the centroid moves in the
gravitational field. Fig. \ref{fig3} shows such a graph. Again as
is expected, increasing $\tau$ produces more spin decoherence  and
the entanglement falls.

\begin{figure}
\begin{center}
\includegraphics[width=8cm,height=8cm,angle=0]{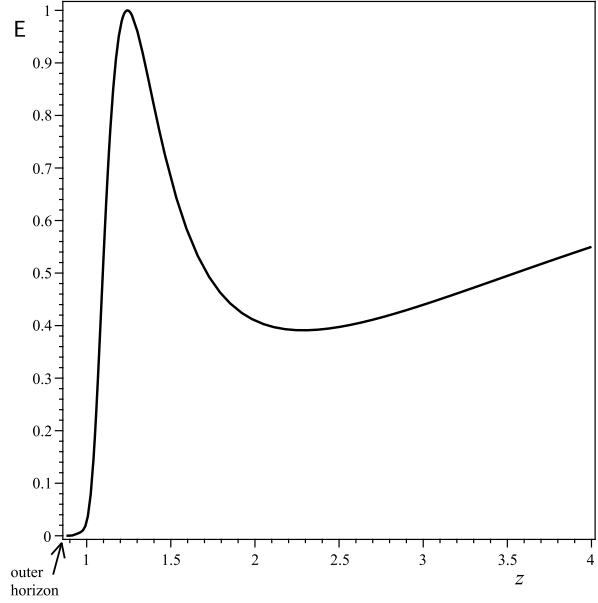}
\end{center}
    \caption{Plot of $\textsf{E}$ for fixed $\xi$, $\tau$, $q$ and $\beta$ but varying $z$.
    Here $\xi$ is chosen such that there exists two event horizons for the spacetime. $z$ begins from the outer
    horizon where $\textsf{E}$ vanishes. There is a peak which
    specifies the radius of a circular orbit on which no
    spin decoherence occurs and the entanglement remains the unity. Also, the curve takes a
    minimum which describes a circular orbit with maximum spin decoherence. The curve
    asymptotically reaches to the unity. }\label{fig4}
\end{figure}

More interesting here is to plot the entanglement in terms of the
radius of the circular orbit characterized by $z$. Fig. \ref{fig4}
shows $\textsf{E}$ versus $z$ for $\xi^2=0.16$, $\beta=1$, $q=0.6$
and $\tau=5\tau_0$. In this case there exist two horizons at
$z=0.2$ and $z=0.8$, however the lower limit of $z$ is taken to be
the outer horizon $z=0.8$. At any horizon $\Theta$ diverges, then
the cosine and sine terms in (\ref{C}), (\ref{S}) oscillate
rapidly yielding a zero value for the integrals. Then the
entanglement (\ref{emp}) is completely destroyed, leading to a
fatal error in quantum communication at the horizon. On the other
hand, the accessible zero of $\Theta$ occurs at $z=1.24$ where the
Wigner rotation (\ref{wigop}) becomes the identity operator. On
such a circle the curvature part and the acceleration part of
(\ref{lamb}) cancel out each other, leading to a zero value for
$\Theta$. Thus, for the corresponding circular orbit $\textsf{E}$
remains the unity, that is, there is no decoherence for the spin
Bell states. This result is independent of values of $q$, $\beta$
and $\tau$. As $z$ grows, the curve takes a minimum at $z=2.25$
which determines a circular orbit on which the effects of
acceleration and curvature add to produce a maximum spin
decoherence. Note that at $z\rightarrow\infty$, both the curvature
part and the acceleration part of (\ref{lamb}) vanish. Hence, as
(\ref{Thetacb}) shows, $\lim_{z\rightarrow\infty}\Theta=0$ and the
curve asymptotically approaches to 1.

\begin{figure}
\begin{center}
\includegraphics[width=8cm,height=8cm,angle=0]{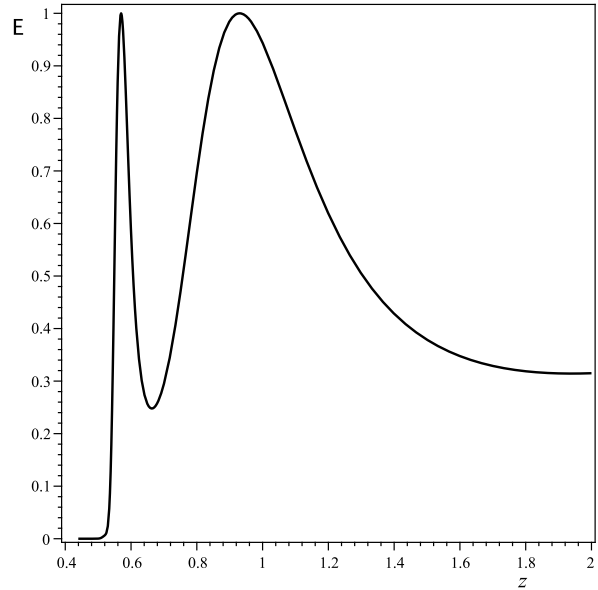}
\end{center}
     \caption{The same plot of figure 4 but for the case of naked singularity. In the vicinity
    of singularity $\textsf{E}$ vanishes. $\xi$ is chosen such that $\Theta$ has two zeros at which the peak
    of curve occurs. The peaks designate two circles on which the entanglement remains intact. Also
    the curve takes two minima indicating two circles of maximum decoherence. The curve asymptotically
    turns to the unity.}\label{fig5}
\end{figure}

\begin{figure}
\begin{center}
\includegraphics[width=8cm,height=8cm,angle=0]{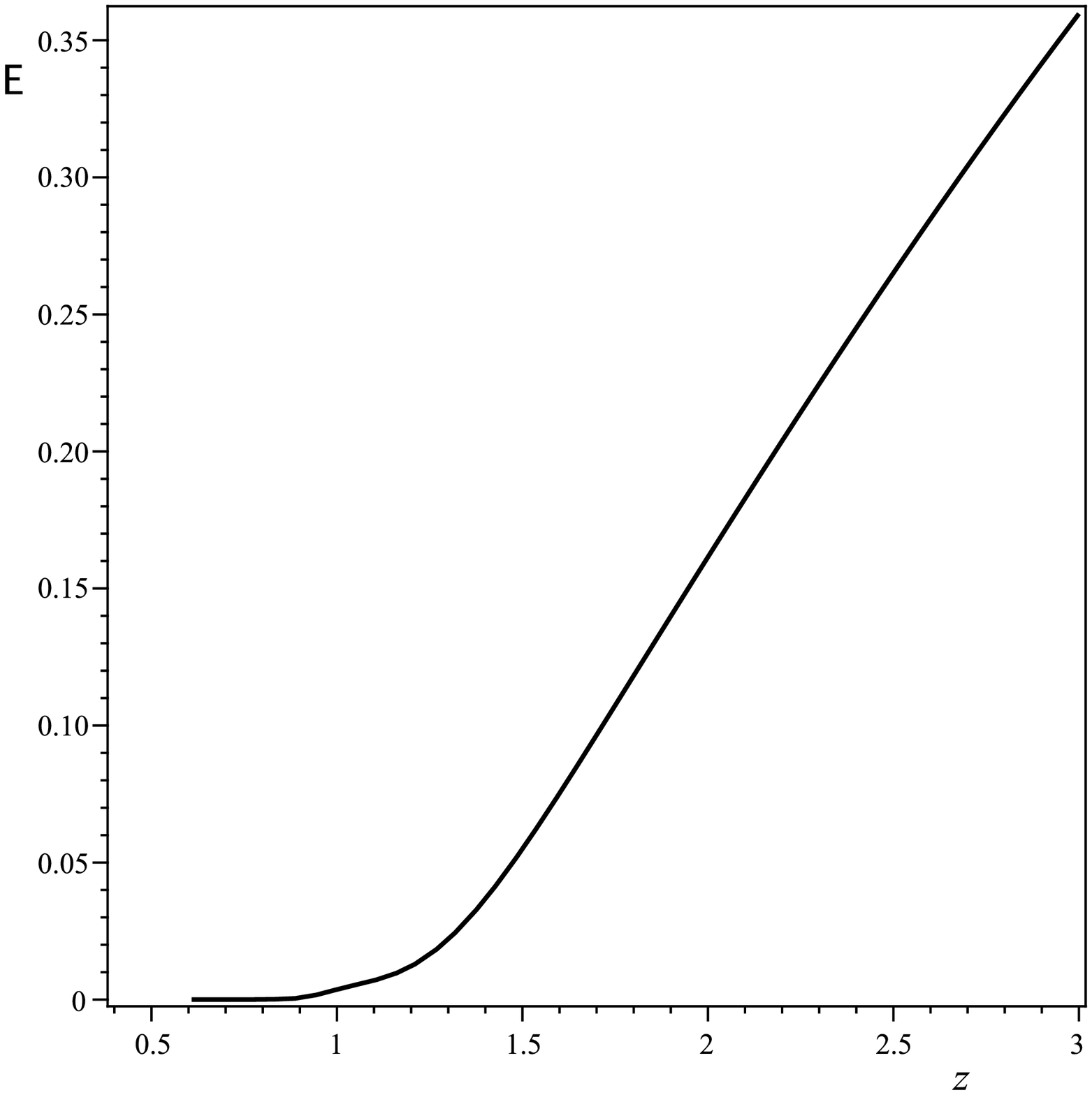}
\end{center}
    \caption{Plot of $\textsf{E}$ versus $q$ for the case of naked singularity, but
    $\xi$ is chosen such that no zeros occur for $\Theta$. Then the curve grows from zero at $z=0$
    and reaches asymptotically to the unity.}\label{fig6}
\end{figure}

In the case of naked singularity $\Theta$ only diverges at $z=0$,
where,  the entanglement vanishes. Fig. \ref{fig5} and Fig.
\ref{fig6} show $\textsf{E}$ in terms of $z$ for this case.
However, the curve in Fig. \ref{fig5} is depicted for
$\xi^2=0.265$ so $\Theta$ has two accessible zeros at $z=0.57$ and
$z=0.93$ where the Wigner rotation becomes the identity operator
and so the entanglement remains intact. There is no spin
decoherence on these two circles, independent of the values of
$q$, $\beta$ and $\tau$. Also the curve takes two minima at
$z=0.65$ and $z=1.96$ indicating two circles of maximum
decoherence. The curve asymptotically turns to the unity. In Fig.
\ref{fig6} the curve is sketched for $\xi^2=0.5$ and then there is
no zeros for $\Theta$. By increasing $z$, the entanglement grows
from zero at $z=0$ and reaches asymptotically to the unity.

That on some circles $\textsf{E}$ remains the unity may suggest a
possibility for a perfect quantum communication around a charged
black hole. The properly entangled spins of the system can
preserve the stored information during the motion on such circles.

\section{Frame transformation}

Consider a local transformation between two instantaneously
coincident inertial frames, as
\begin{equation}\label{tettran}
      {e_a}^\mu\longrightarrow
      {\tilde{e}_a}^{\,\,\,\mu}={\mathcal{T}_a}^b(x){e_b}^\mu,
\end{equation}
where
${\mathcal{T}_a}^c(x){\mathcal{T}_b}^d(x)\eta_{cd}=\eta_{ab}$.
Notice that the definition (\ref{tet}) remains intact under this
tetrad transformation which can arise from a coordinate
transformation $x^\mu\longrightarrow\tilde{x}^\mu$. However, as
(\ref{lamb}) implies, the new inertial observer who uses
${\tilde{e}_a}^{\,\,\,\mu}$  obtains different values for the
components ${\lambda^a}_b(x)$, and regarding (\ref{dabelu}), we
recognize that the Wigner rotation is not invariant under the
transformation (\ref{tettran}). Consequently, the spin
entanglement is not frame independent. To illustrate the
situation, let us simply consider the Schwarzschild black hole by
letting $e^{A(r)}=e^{-B(r)}=\sqrt{1-\frac{r_s}{r}}$ in the metric
(\ref{metric}). Consequently, the tetrad (\ref{dstet}) reduces to
\begin{equation}\label{sctet}
      {e_0}^t=\frac{1}{c\sqrt{1-\frac{r_s}{r}}},\,\,\,\,\,\,\,\,\,\,
      {e_1}^r=\frac{1}{\sqrt{1-\frac{r_s}{r}}},\,\,\,\,\,\,\,\,\,\,
      {e_2}^{\theta}=\frac{1}{r},\,\,\,\,\,\,\,\,\,\,{e_3}^{\phi}=\frac{1}{r\sin\theta},
\end{equation}
which is used by a static local observer. Note that this is
singular at the horizon $r=r_s$. Then, the local Wigner rotation
elements (\ref{16}) reduce to
\begin{eqnarray}\label{wsc}
     {w^1}_3=-{w^3}_1=c\left(\frac{1-\frac{3r_s}{2r}}{r\sqrt{1-\frac{r_s}{r}}}\right)
     q\sqrt{q^2+1}\left(\sqrt{q^2+1}
     -\frac{q\,p}{\sqrt{p^2+1}+1}\right),
\end{eqnarray}
which can be used for finding the elements of the Wigner rotation
operator and then calculating the entanglement (\ref{emp}). These
elements diverge at the horizon and so the entanglement will be
completely destroyed there, leading to a fatal error in quantum
communication. Also they vanish at the circle $r=\frac{3r_s}{2}$
where the entanglement will remain intact. Now, let us transform
the present spherical coordinates $(t,r,\theta,\phi)$ to, for
instance, the Kruskal coordinates $(T,R,\theta, \phi)$ via the
relations
\begin{equation}
       R^2-c^2T^2=4r_s^2\left(\frac{r}{r_s}-1\right)e^{\frac{r}{r_s}},\,\,\,\,\,\,\,\,\,\,\,\,\,
       \frac{cT}{R}=\tanh\left(\frac{ct}{2r_s}\right).
\end{equation}
In the Kruskal coordinates the Schwarzschild metric becomes
\begin{equation}
       ds^2=\frac{r_s}{r}e^{-\frac{r}{r_s}}\left(-c^2dT^2+dR^2\right)+r^2\left(d{\theta}^2+\sin^2\theta\,d\phi^2\right),
\end{equation}
where the radial coordinate $r$ is now interpreted as a function
of $T$ and $R$. Accordingly, we choose a new tetrad
${\tilde{e}_a}^{\,\,\,\mu}$ as
\begin{equation}\label{ktet}
      {\tilde{e}_0}^{\,\,\,T}=\frac{1}{c}\sqrt{\frac{r}{r_s}}\,e^{\frac{r}{2r_s}},\,\,\,\,\,\,\,\,
       {\tilde{e}_1}^{\,\,\,R}=\sqrt{\frac{r}{r_s}}\,e^{\frac{r}{2r_s}},\,\,\,\,\,\,\,
      {\tilde{e}_2}^{\,\,\,\theta}=\frac{1}{r},\,\,\,\,\,\,\,\,{\tilde{e}_3}^{\,\,\,\phi}=\frac{1}{r\sin\theta},
\end{equation}
which is not singular at the horizon. According to
(\ref{tettran}), we find that the this tetrad is related to the
static tetrad (\ref{sctet}) by a local transformation
\begin{equation}
      {\mathcal{T}_a}^c(x)=\left(
       \begin{array}{cccc}
      \frac{1}{2}\sqrt{\frac{e^{-\frac{r}{r_s}}}{r_s(r-r_s)}}\,R
      &-\frac{c}{2}\sqrt{\frac{e^{-\frac{r}{r_s}}}{r_s(r-r_s)}}\,T&
      0&0\cr & & &\cr-\frac{c}{2}\sqrt{\frac{e^{-\frac{r}{r_s}}}{r_s(r-r_s)}}\,T
      &\frac{1}{2}\sqrt{\frac{e^{-\frac{r}{r_s}}}{r_s(r-r_s)}}\,R&0&0\cr
       & & &\cr0&0&1&0\cr
      0&0&0&1
       \end{array}
      \right).
\end{equation}
This transformation resembles a Lorentz boost along the 1-axis and
so the new local inertial observer falls into the black hole when
$T>0$. Now, repeating the steps of Section 2 with the tetrad
(\ref{ktet}), we obtain, instead of (\ref{wsc}), the local Wigner
rotation elements
\begin{equation}\label{wk}
       {\tilde{w}^1}_{\,\,\,3}=-{\tilde{w}^3}_{\,\,\,1}=\frac{c}{4r}\sqrt{\frac{e^{-\frac{r}{r_s}}}{rr_s}}\left(3+\frac{r}{r_s}\right)
     \tilde{q}\sqrt{\tilde{q}^2+1}\left(\sqrt{\tilde{q}^2+1}
     -\frac{\tilde{q}\,\tilde{p}}{\sqrt{\tilde{p}^2+1}+1}\right),
\end{equation}
where $\tilde{q}$ and $\tilde{p}$ relate to the four-momentum of
the centroid and the four-momentum of particles, respectively,
that are chosen by the falling observer. Since these elements are
not singular at the horizon, the spin entanglement can have a
nonzero value and falling observers just on the horizon can
perform a quantum information task with a finite error.
Furthermore, they can do this beyond the horizon $r<r_s$, until
the physical singularity $r=0$. Therefore, the features of quantum
communication in a curved spacetime using the entangled spins
depend on the tetrad that is used by the observer.

\section{Conclusion}

In this paper we discussed the effect of a gravitational field on
the spin entanglement of a system of two spin-$\frac{1}{2}$
particles moving in the gravitational field. We described the
system by a two-particle Gaussian wave packet represented in the
momentum space with a definite width around the momentum of a mean
centroid. We considered the mean centroid for describing the
motion of the system in the gravitational field which was a
classic field in our approach. Both the acceleration of the
centroid and the curvature of the gravitational field caused to
produce a Wigner rotation acting on the wave packet. We confined
the argument to spherically symmetric and static gravitational
field such as a charged black hole. At a given initial point in
the curved spacetime, the momenta of the particles were taken to
be separable, while the spin part was chosen to be entangled as
the Bell states. By calculating the reduced density operator, we
focused on the spin entanglement of the system.

For any spherically symmetric and static spacetime, when the
system, no matter of spin and momentum entanglements, falls along
a radial geodesic, the Wigner rotation is trivial and so the spin
density matrix remains intact. Moreover, for a charged black hole,
depending on the mass and the charge of black hole, there can be
circular paths with determined radii on which the spin
entanglement is invariant. For instance, it always remains the
unity, provided spins be entangled initially as one of the Bell
states. This result is a consequence of cancellation of the
acceleration and the curvature on such circles. The spin
entanglement descends to zero by increasing  the angular velocity
of the centroid as well increasing the proper time during which
the centroid moves on its circular path around the center of
charged black hole.

It is remarkable that the features of quantum communication in a
curved spacetime using the entangled spins depend on the tetrad
that is used by the observer.

As an extension of this work, one can provide the initial state
with a specified spin entanglement and also a specified momentum
entanglement, then discuss the final spin entanglement, as well as
the final momentum entanglement.

\end{spacing}
\end{document}